\documentclass[epj]{webofc}
\usepackage[utf8]{inputenc}
\usepackage[varg]{txfonts}   % Web of Conferences font
\usepackage{JackCommands}
\usepackage{booktabs}
\usepackage{xcolor}
\usepackage{amssymb}
\usepackage{geometry}
\usepackage{ragged2e}
\usepackage{graphicx}%graphics
\usepackage{slashed} %feynman slash notation
\usepackage{array} %additional array options
\usepackage{braket}
\usepackage{simplewick}
\usepackage{commath}
\usepackage{tabularx}
\usepackage{booktabs}
\usepackage{cancel}
\usepackage{tikz}

\definecolor{darkred}{rgb}{0.4,0.0,0.0}
\definecolor{darkgreen}{rgb}{0.0,0.4,0.0}
\definecolor{darkblue}{rgb}{0.0,0.0,0.4}
\usepackage[bookmarks,linktocpage,colorlinks,
    linkcolor = darkred,
    urlcolor  = darkblue,
    citecolor = darkgreen]{hyperref}
\wocname{EPJ Web of Conferences}
\woctitle{lattice2017}
\begin{document}
%%%%%%%%%%%%%%%%%%%%%%%%%%%%%%%%%%%%%%%%%%%%%%%%%%%%%%%%%%%%%%%%%%%%%%%%%%%%%
%
\selectlanguage{english}
%----------------------------------------------------------------------------
\title{%
Electric Dipole Moment Results from lattice QCD
}
%----------------------------------------------------------------------------
\author{%
\firstname{Jack} \lastname{Dragos}\inst{1}\fnsep\thanks{\email{dragos@frib.msu.edu}} \and
\firstname{Thomas}  \lastname{Luu}\inst{2} \and
\firstname{Andrea} \lastname{Shindler}\inst{1} \and
\firstname{Jordy}  \lastname{de Vries}\inst{3}
% etc.
}
%----------------------------------------------------------------------------
\institute{%
  Facility for Rare Isotope Beams, Physics Department, Michigan State University, East Lansing, Michigan.
  \and
  Institute for Advanced Simulation (IAS-4), Institut f\"ur Kernphysik (IKP-3), and J\"ulich Center for Hadron Physics, FZJ.
  \and
  Nikhef, Amsterdam, the Netherlands.
}
%----------------------------------------------------------------------------
\abstract{%
  We utilize the gradient flow to define and calculate electric dipole moments induced by the
  strong QCD $\theta$-term and the dimension-6 Weinberg operator.
  The gradient flow is a promising tool to simplify the renormalization pattern of local operators.
    The results of the nucleon electric dipole moments are calculated on PACS-CS gauge fields (available from the ILDG)
  using $N_{f}=2+1$, of discrete size $32^{3}\times 64$ and spacing $a \simeq 0.09$ fm.
  These gauge fields use a renormalization-group improved gauge action and
  a non-perturbatively $O(a)$ improved clover quark action at $\beta = 1.90$, with $c_{SW} = 1.715$.
  The calculation is performed at pion masses of $m_{\pi} \simeq 411,701$ MeV.
}
%----------------------------------------------------------------------------
\maketitle
%----------------------------------------------------------------------------
\section{Introduction}\label{intro}

%% Understanding strong CP violation effects is currently achieved by analysing the CP violation coming from the neutron (and Proton) Electric Dipole Moment (pEDM & nEDM).

The study of the electric dipole moment of the neutron and proton (nEDM \& pEDM)
provides us a tool for understanding how different sources of CP violation manifest in hadronic systems.
Experimentally, the total nEDM has been bounded by \(|d_{n}| < 3.0 \times 10^{-13} \)~e~fm \cite{Olive:2016xmw}.
A weaker bound of \(|d_{p}|< 2.0 \times 10^{-12} \)~e~fm \cite{Graner:2016ses}
has been achieved indirectly from the limit of the Hg EDM, in this way overcoming the difficulties of observing an EDM of a charged system.

The Standard Model (SM) contains the ``\(\theta\)-term'' which
can induce an EDM in the neutron and proton. At the same time,
Beyond the Standard Model (BSM) physics can also provide contributions to EDMs. At hadronic energies these can be parametrized by effective higher-dimensional operators that violate CP.

Due to recent advancements in computational power and theoretical developments,
lattice QCD is fast approaching the precision needed to calculate the EDM of the neutron and proton \cite{Alexandrou:2015spa,Guo:2015tla}.
The key feature when providing this \textit{ab-initio} calculation from lattice QCD,
is that the individual contributions from the \(\theta\)-term and all the BSM operators
can be calculated individually. These calculations, combined with future experimental
results, can provide unique constraints for the different types of BSM
CP violations responsible for a non-vanishing EDM.

In this work, we analyze the nEDM and pEDM induced by
the \(\theta\) and Weinberg terms individually.
We use the gradient flow to define the $\theta$-term and the Weinberg operator on the lattice.
For the $\theta$-term, the use of gradient flow circumvents the problems associated with renormalization.
For the Weinberg operator it provides a powerful method to connect
the divergent-free definition of the corresponding correlation functions at non-vanishing flow time
with the physical matrix element.

\section{Theory}\label{theory}

The QCD Lagrangian in Euclidean space without strong CP violation, has the form

\be
\mathcal{L}_{QCD} = \frac{1}{4}\Gmunu^{a}\Gmunu^{a} + \sum_{q=u,d,s} \overline{\psi}_{q} \left(\gamma_\mu D_{\mu} + m_{q}\right)\psi_{q}\,,
\ee
where \(\Gmunu^{a}\) denotes the gluonic field strength,
$\psi_{u,d,s}$ denote the up, down, and strange quarks,
 \(\gamma_\mu D_{\mu}\) the gauge-covariant derivative, and \(m_{q}\) the fermion quark masses .
We consider two CP-violating terms in this contribution. The first is
the ``\textit{\(\theta\)-term}'' proportional to the topological charge density $q(x)$
\bes
-i\theta q(x) \equiv -i\theta \frac{1}{32\pi^2}\epsilon_{\mu\nu\rho\sigma}\Trace\left[\Gmunu(x)G_{\rho\sigma}(x)\right],
\ees
and the second is the Weinberg operator \cite{Weinberg:1989dx}
\bes
- i \frac{\alpha_{\widetilde{G}}}{\Lambda^{2}}\mathcal{O}_{W}(x)
\equiv - i \frac{\alpha_{\widetilde{G}}}{\Lambda^{2}}\,\frac{1}{3}f^{ABC} \widetilde{G}^{A}_{\mu\nu}(x) G^{B}_{\mu\rho}(x) G^{C}_{\nu\rho}(x).
\ees
\(\theta\) and \(\alpha_{\widetilde{G}}\) are the coupling coefficients
of the topological charge density and Weinberg operator respectively.
The Weinberg operator is an effective operator of dimension 6 and therefore suppressed by two powers of the unknown high-energy matching scale $\Lambda$, where the Weinberg operator is induced.

\section{Lattice Parameters}\label{LatParams}

We performed calculations on the publicly available PACS-CS gauge fields \cite{Aoki:2010wm}
available through the ILDG \cite{BECKETT20111208}. They provide \(N_{f}=2+1\) dynamical-QCD gauge fields,
generated utilizing a non-perturbatively \(O(a)\)-improved Wilson fermion action (\(c_{SW} = 1.715\)) along with an Iwasaki gauge action.
The size of the utilized gauge fields in this paper are \(32^{3}\times 64\)
discrete space-time lattices, with a lattice spacing of \(a \simeq 0.09\) fm (\(\beta = 1.90\)) with \(L = 2.91\) fm.

%% The Markov chain of configurations supplied via the ILDG is incremented by 10 updates to ensure the autocorrelation effects are negligible for nucleon observables.

The current state of the calculation has measurements of the EDM for the neutron and proton
at \(m_{\pi} = 411,701\) MeV (giving \(m_{\pi}L = 5.65,10.32\) respectively).
For all the plots shown in this paper, results for \(m_{\pi}=411\) MeV and \(m_{\pi}=\)701 MeV
are plotted in, respectively, blue and red unless otherwise indicated.

A Gaussian gauge-invariant smearing \cite{Gusken:1989qx}
is applied to the source and sink propagators
used in the construction of the two- and three- point correlation functions.
The smearing fraction \(\alpha =0.7\) was selected when applying the $64$
iterations of the smearing algorithm to both the source and sink.
This provides a root-mean-square (rms) radius of \(r_{rms} = 0.431\) fm which is \(15\%\)
of the spatial extent of the lattice \(L\).

For the vector form factors, we use the renormalization of \(Z_{V} = 0.7354\) taken from \cite{Aoki:2010wm}.

\section{Topological and Weinberg Susceptibilities }\label{Susceptibility}

%We define the topological susceptibility and an analogous {\it Weinberg susceptibility}
%\be
%\chi_{t} = \frac{1}{V} \int d^{4}x d^{4}y \braket{q(x)q(y)}
%\quad , \quad
%\chi_{W} = \frac{1}{V} \int d^{4}x d^{4}y \braket{\mathcal{O}_{W}(x)W(y)},
%\ee
%where \(V\) is the total volume of the lattice. Equivalently
%\be
%\chi_{t} = \frac{1}{V} <Q_{t}^{2}>
%\quad , \quad
%\chi_{W} = \frac{1}{V} <W^{2}>,
%\ee
%where we have defined the topological charge, and the analogous Weinberg charge as the 4-D integral over the lattice:
%\be
%Q_{t} = \int d^{4}x\, q(x) \quad , \quad W = \int d^{4}x\, \mathcal{O}_{W}(x).
%\ee

%Although the gradient flow has been usefull for many quantities in lattice QCD
%\cite{Luscher:2013cpa,Luscher:2010iy,Borsanyi:2012zs,Shindler:2013bia,Fritzsch:2013je,Shindler:2014oha}
We define the topological susceptibility and an analogous {\it Weinberg susceptibility}
using the gradient flow~\cite{Luscher:2011bx,Bruno:2014ova,Chowdhury:2013mea}
with gauge fields defined at non-vanishing flow time \(t_{f}\). The topological susceptibility defined
in this way is finite and free from renormalization ambiguities~\cite{Luscher:2010iy,Borsanyi:2012zs}.
The Weinberg susceptibility is also finite at finite flow time, but contrary to the topological susceptibility,
needs to be connected to the physical observable at vanishing flow time.

%\subsection{Gradient Flow Formalism for Gauge Fields}\label{GradFlow}

%The gradient flow for gauge and fermion fields \cite{Luscher:2011bx} has shown
%to be a powerful tool dealing with discretisation and renormalization difficulties from lattice QCD,
%as it removes the ultra-violate divergences of the theory in terms of a continuous parameter \(t_{f}\).

%The gradient flowed gauge field \(B_{\mu}(x,t_{f})\) is the solution to the differential equation of the form
%\be\label{GradFlowEq}
%\partial_{t_{f}}B_{\mu} = D_{\nu} G_{\mu \nu},
%\ee
%where

%\be

%D_{\nu} \equiv \partial_{\nu} + \left[B_{\nu},\ \cdot\  \right] \quad , \quad
%G_{\mu \nu} \equiv \partial_{\mu}B_{\nu} - \partial_{\nu}B_{\mu } + \left[ B_{ \mu} ,B_{\nu } \right].
%\ee

%The initial condition is selected as the regular gauge field \(B_{\mu}(x,0) =A_{\mu}(x)\). In other words, \(B_{\mu}\) is flowed from \(A_{\mu}\), and the differential equation (\ref{GradFlowEq}) dictates how to evolve over \(t_{f}\).

\subsection{Susceptibility results}\label{SuscRes}

Expressed in terms of the flow time, the susceptibilities read
\be
\chi_Q(t_{f}) = \frac{1}{V} \int d^{4}x d^{4}y \braket{q(x,t_{f})q(y,t_{f})} = \frac{1}{V} <Q(t_{f})^{2}>
\ee
\be
\chi_{W}(t_{f}) = \frac{1}{V} \int d^{4}x d^{4}y \braket{\mathcal{O}_{W}(x,t_{f})\mathcal{O}_{W}(y,t_{f})} = \frac{1}{V} <\mathcal{O}_{W}(t_{f})^{2}>
\ee

\begin{figure}
\vspace{-1cm}
  \includegraphics[trim={0cm 0cm 5mm 0cm},width=0.5\textwidth]{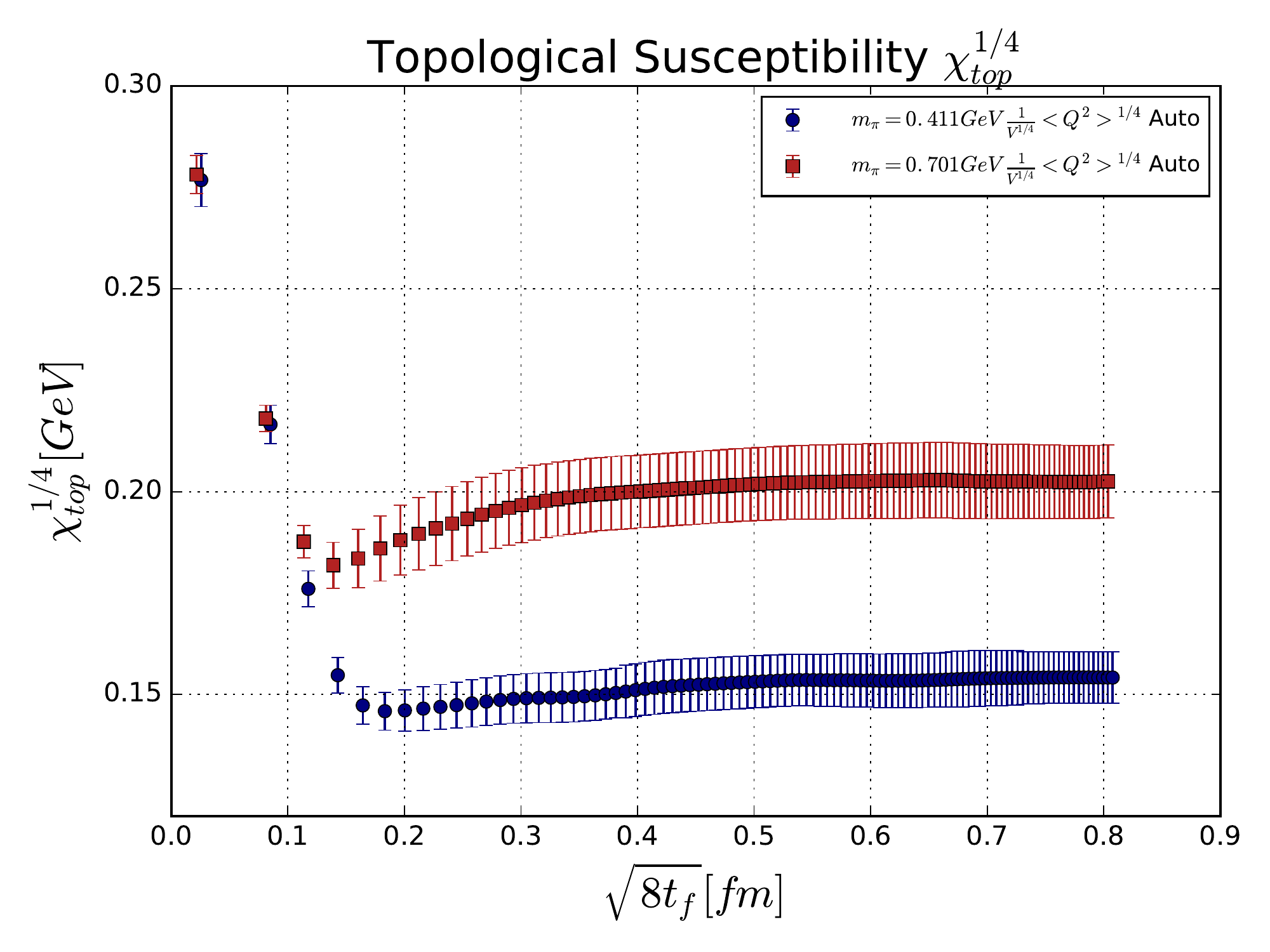}
  \includegraphics[trim={0cm 0cm 5mm 0cm},width=0.5\textwidth]{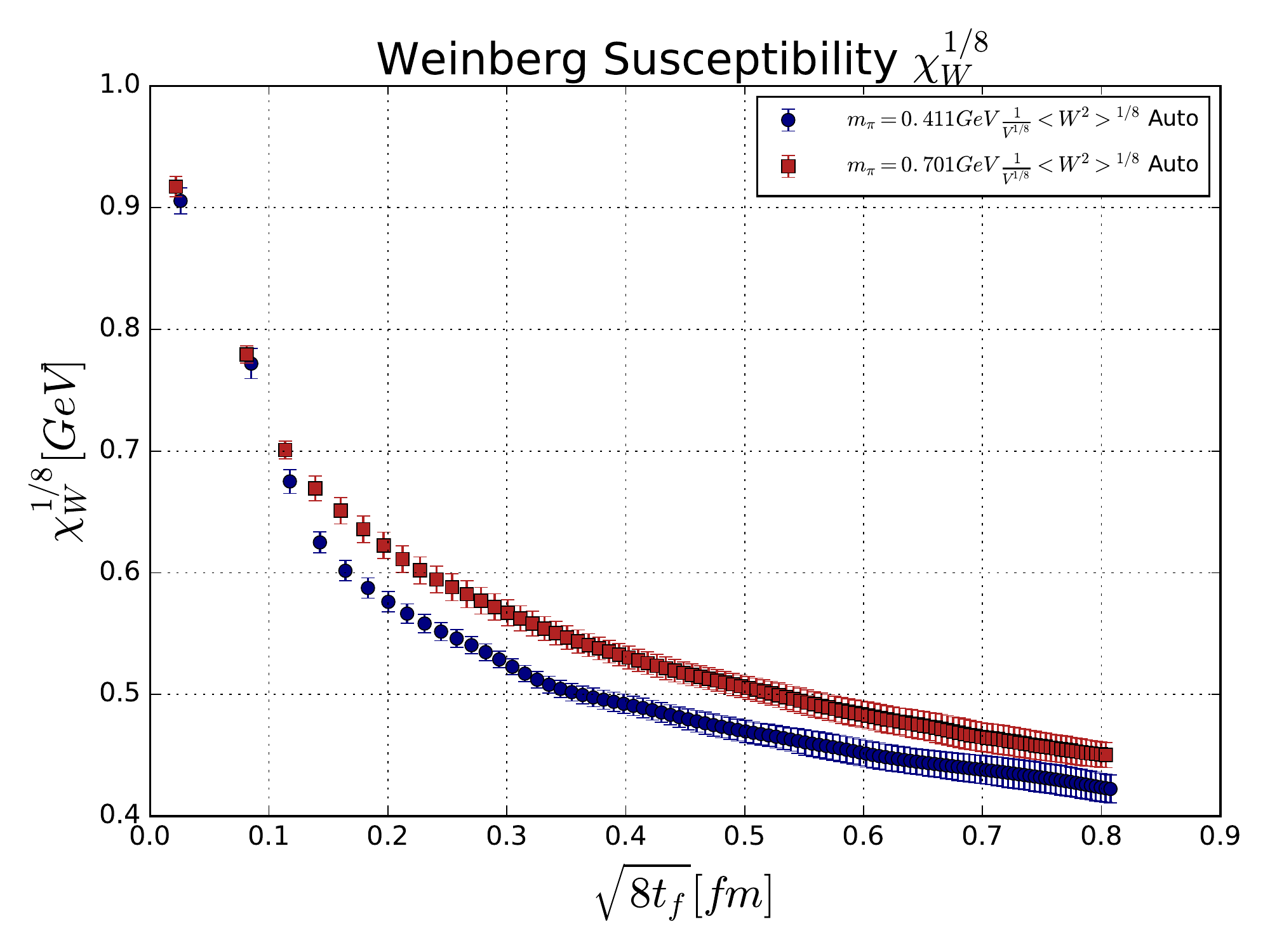}
  \caption{\label{ChitRes} Topological (left) and Weinberg (right) susceptibilities in GeV,
    plotted against the flow-time radius $\sqrt{8t_f}$.
    Blue and red points are the \(m_{\pi} = 411, 701\) MeV results respectively.
    Errors were estimated using the autocorrelation analysis technique described in \cite{Wolff:2003sm}.}
\end{figure}
\begin{figure}
\vspace{-0.5cm}
\centering
  \includegraphics[trim={0cm 0cm 5mm 0cm},width=0.5\textwidth]{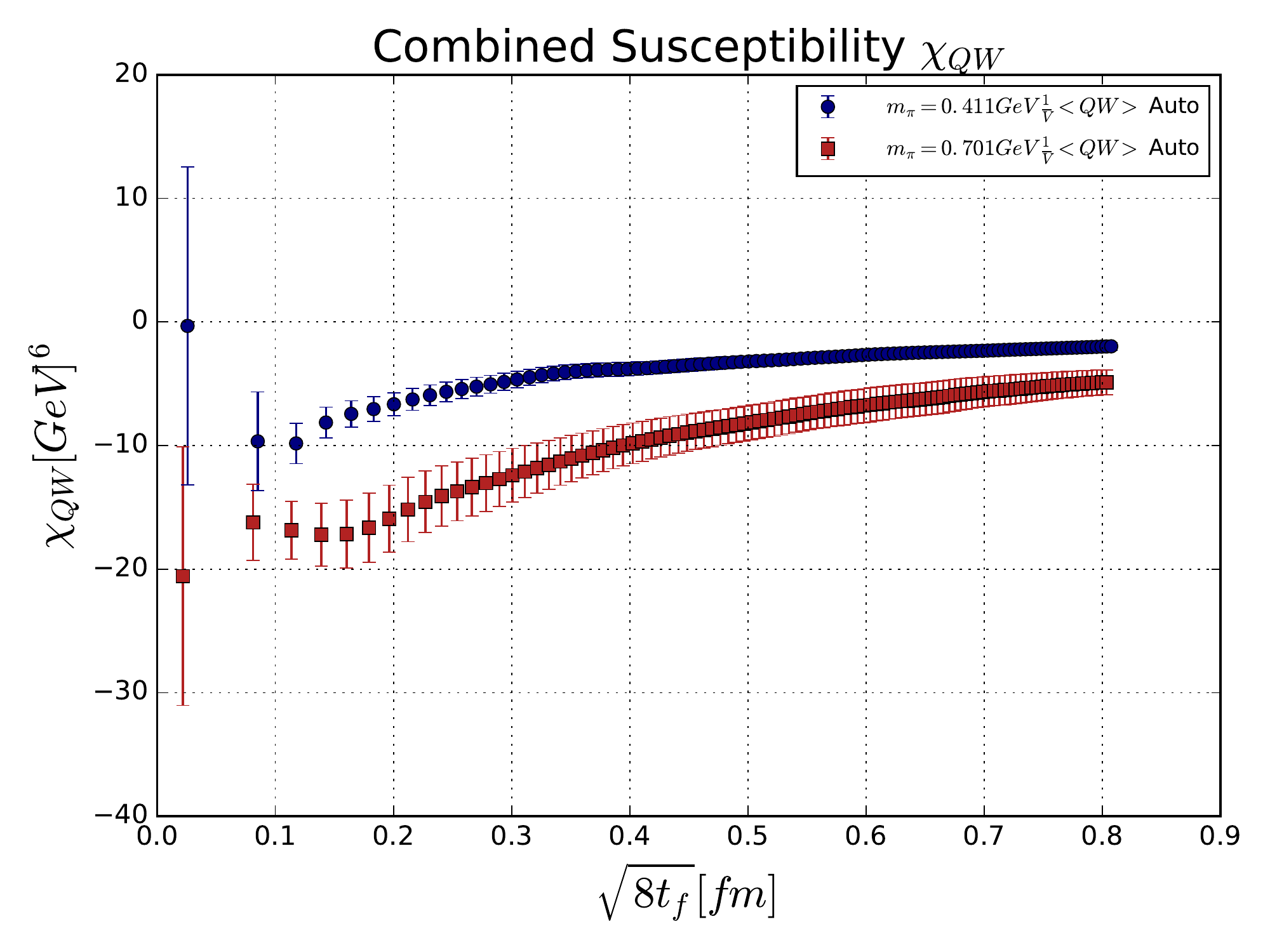}
  \caption{\label{ChitQWRes} Cross correlation \(\braket{Q\mathcal{O}_{W}}\) in GeV\(^{6}\) against the flow-time radius $\sqrt{8t_f}$.
    Blue and red points are the \(m_{\pi} = 411, 701\) MeV results respectively.
    Errors were estimated using an autocorrelation analysis technique described in \cite{Wolff:2003sm}.}
\end{figure}
For the topological susceptibility results in the left of fig.\ref{ChitRes},
at small flow-time radius (ftr) \(\sqrt{8t_{f}}\) we observe the expected short-distance singularities, while
as we approach ftr of the order of the lattice spacing \(\sqrt{8t_{f}} \approx 0.1\) fm,
we see the divergences already being suppressed,
but discretisation effects causing a ``dip'' in the flow time dependence.
Once the ftr is of the order of $4-5$ times the lattice spacing, \(\sqrt{8t_{f}} \approx 0.4\) fm,
as expected no more flow time dependence is observed for this quantity.
The result can simply be read off from any of \(\sqrt{8t_{f}} > 0.4\) fm,
as the results are constant and nearly identical in this range.
Any residual flow-time dependence is just a lattice artifact and will vanish in the continuum limit.

The Weinberg susceptibility in the right plot of fig.\ref{ChitRes}
also shows an indication of possible divergences at small \(\sqrt{8t_{f}}\),
%as well as some slight discretisation effects at \(\sqrt{8t_{f}} \approx 0.1 fm\).
but unlike the topological susceptibility, we observe a non-trivial flow-time dependence over the whole
range of \(\sqrt{8t_{f}}\). The analysis of the flow-time dependence for the Weinberg susceptibility is ongoing.

A potentially interesting quantity to analyze, to understand the flow-time dependence of the Weinberg operator,
is the \textit{``cross correlation''} between the topological charge \(Q\)
and the integrated Weinberg operator \(\mathcal{O}_{W}\) shown in fig.\ref{ChitQWRes}.
On general grounds the Weinberg operator mixes under renormalization with the
topological charge density~\cite{Bhattacharya:2015rsa} and such correlator can provide a useful
tool to disentangle contributions to the flow-time dependence of the Weinberg operator itself.

%Since \(Q_{t}\) has no flow time dependence in \(\sqrt{8t_{f}} > 0.4 fm\),
%the only flow time dependence present is the one due to a single power of \(W\).
%This might prove beneficial, as the flow time dependence is less pronounced compared
%to the Weinberg susceptibility. The presences of a negative result between
%the expectation value \(\braket{Q(t_{f})W(t_{f})}\) is a curious find,
%as the total result is imaginary when computed in units of \(GeV\).

\section{Nucleon Observables in CP-breaking theory}\label{NuclObsCP}

In this section, we show how to calculate nucleon observables in a
CP-violating vacuum from lattice QCD.
Along with this, we show the preliminary results demonstrating this procedure
for both the \(\theta\) and Weinberg CP-violating contributions
to the nucleon observables.
The method for the \(\theta\)-term has been discussed in detail in \cite{Shindler:2015aqa}.

To study nucleon observables in a CP-violating vacuum we use a perturbative approach treating
every CP-violating operator as insertion in the nucleon correlation functions evaluated
in the usual QCD CP-conserving vacuum~\cite{Shintani:2005xg,Shindler:2015aqa}.

\subsection{Nucleon Mixing Angle}\label{NucMixAng}

To understand how the nucleon reacts to a theory that includes CP-violation,
we can calculate the nucleon CP-violating mixing angle, $\alpha_N$,
which tells us how much the nucleon spinor is modified when placed in the CP-violating vacuum~\cite{Shintani:2005xg,Shindler:2015aqa}:
\be
u^{\cancel{CP}}_{N}(\p,s) = e^{i\alpha_{N}\gamma_{5}} u_{N}(\p,s).
\ee
%
%Since \(\alpha_{N}^{\underline{p}}\) is CP-odd, the expansion in \(\underline{p}\) contains only odd powers:
%
%\be
%\alpha_{N} (\underline{p}) =  \underline{p} \alpha^{(1)}_{N} + \Ord(\underline{p}^{3}).
%\ee
%$\alpha_N$ depends on the CP-violating source, but for simplicity we omit to label it. The CP-vilating sour
The leading contribution of a generic CP-violating local operator $\mathcal{O}$ evaluated at flow time $t_f$,
to the mixing angle $\alpha_N^{(1)}$, can be estimated computing the ratio between
\be
\GOtwozt = \sumx  \Trace \left\lbrace  \Gamma \gamma_{5}  \braket{\chixt \chizerobar \mathcal{O}(t_{f})}\right\rbrace\,,
\ee
and the standard nucleon two-point correlation function
\be\label{AlphaRat}
\frac{\GOtwoztGfour}{\GtwoztGfour} \xrightarrow{t \gg 0} \alpha^{(1)}_{N}(t_f).
\ee
%are used in a CP-conserving and CP-violating vacuum.
%The standard two-point correlation function is defined as:
%
%\be\label{G2ptcorr}
%\Gtwopt = \sumx \Fpx \Trace \left\lbrace  \Gamma \braket{ \chixt \chizerobar} \right\rbrace,
%\ee
where \(\chixt\) is an interpolating operator with the quantum numbers of a nucleon (proton and neutron in \(N_{f} = 2+1\)),
\(\Gamma\) is used in the trace to project out specific spin combinations, and a Fourier transform from position  to momentum space is used to analyze a nucleon of momentum \(\p\).
In our case, the CP-violating operator $\mathcal{O}$ is the topological charge and the integrated Weinberg operator.
The flow time dependence needs to be analyzed in addition to the ground state
saturation through a large source-sink separation \(t\gg 0\).

\subsection{CP-odd Form Factor \(F_{3}\) and the Electric Dipole Moment}\label{EDMRes}
\begin{figure}
\vspace{-1cm}

  \includegraphics[trim={0cm 0cm 0.5cm 0cm},width=0.5\textwidth]{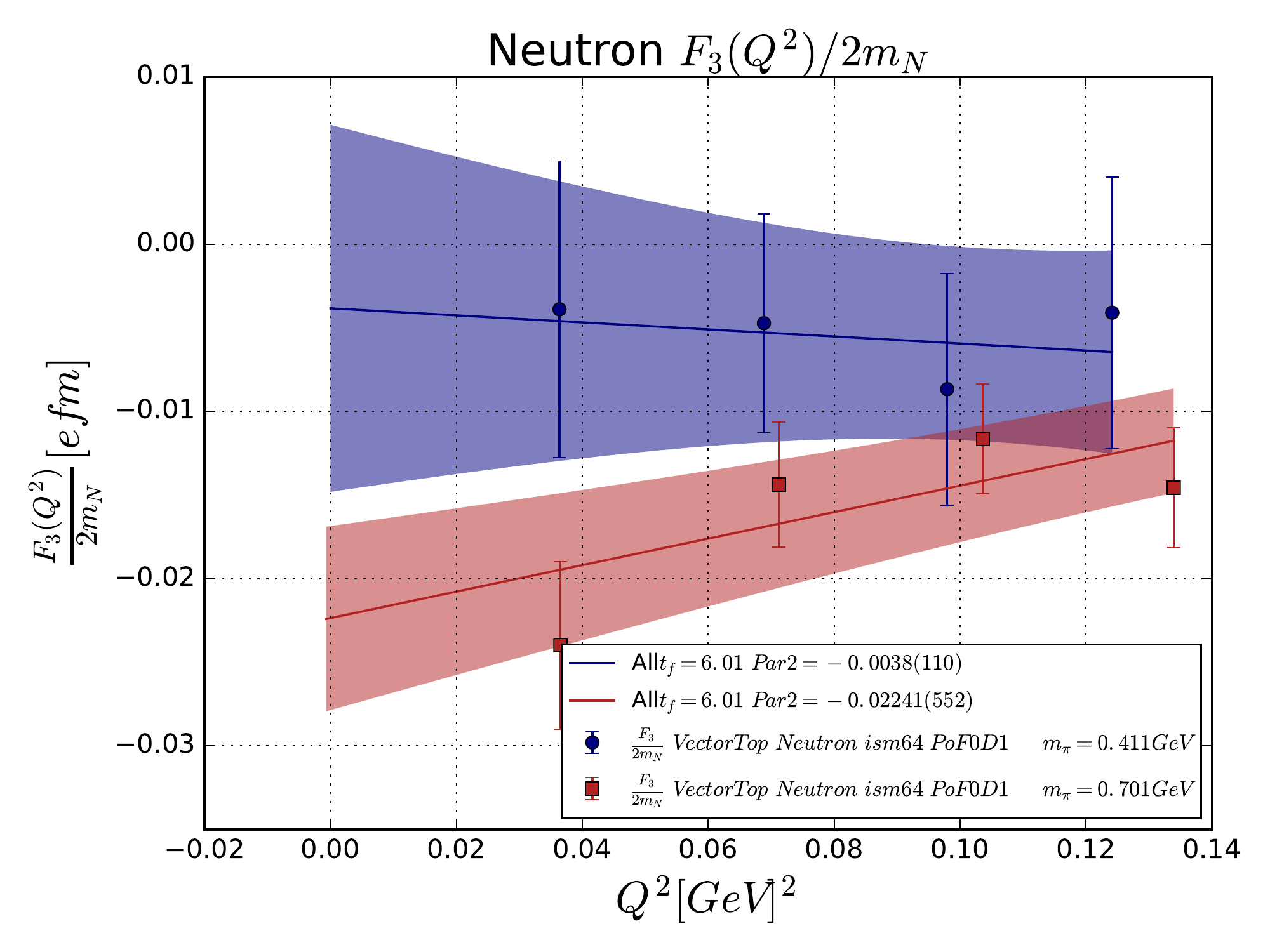}
  \includegraphics[trim={0cm 0cm 0.5cm 0cm},width=0.5\textwidth]{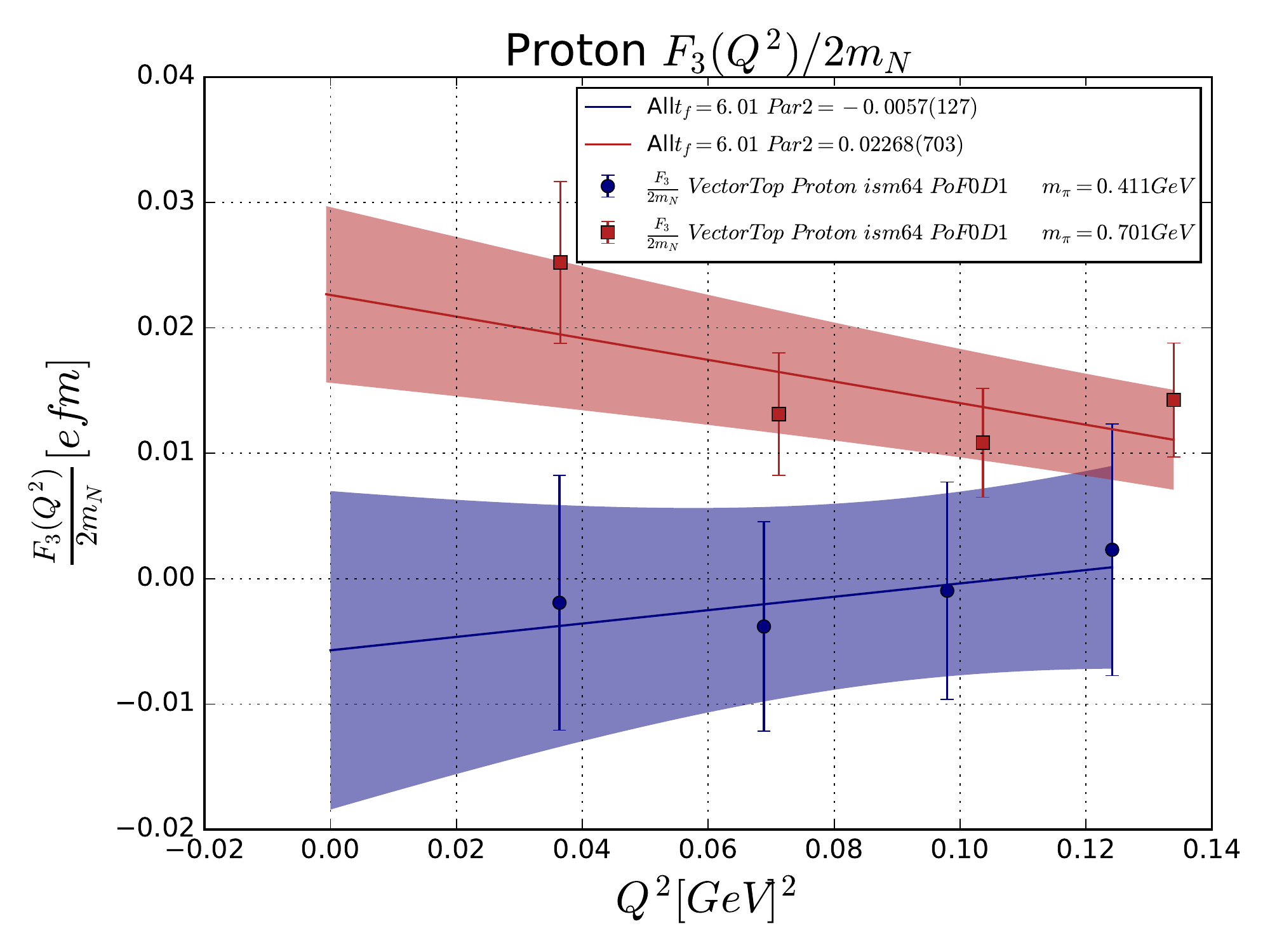}
  \caption{\label{TopF3} CP-odd vector form factor \(F_{3}/2m_N\) in units of \(\theta\) for the neutron (left) and proton (right),
    induced via the \(\theta\)-term CP-violating vacuum, plotted against transfer momentum \(Q^{2}\).
    Blue and red points are the \(m_{\pi} = 411, 701\) MeV results respectively.
    The linear fit to \(Q^{2}\rightarrow 0\) is used to extract the value for the neutron and proton EDM \(d_{n/p}\).}
\end{figure}
The CP-odd form factor for the nucleon \(F_{3}(Q^{2})\), which is related to to the
nucleon EDM via $d_{n/p}=F_3(0)/2m_N$, is only present in a theory that breaks CP-symmetry.
Thus, to access this form factor from lattice QCD, we need to look at observables corresponding to form factors,
in a CP-violating vacuum.
%\be\label{F3TOEDM}
%\frac{F^{P/N}_{3}(Q^{2})}{2M_{N}} \xrightarrow{small\ Q^{2}} d_{P/N} + S_{P/N} Q^{2} + \Ord (Q^{4}).
%\ee
% \subsection{Results}\label{results}
In addition to standard two- and three- point correlation functions a modified three-point correlation
of the form
\be
G^{(\mathcal{O})}_{3} ( \Gamma ;  \pp , t ; \q, \tau, t_{f} )
 = \sumxy \Fppx \Fqy \Trace \left\lbrace \Gamma \braket{ N(\x,t) \mathcal{J}_{\mu}(\y,\tau) \overline{N}(0) \mathcal{O}(t_{f}) } \right\rbrace,
\ee
is required to gain access to the CP-violating form factor \(F_{3}(Q^{2})\) arising
from a (flowed) CP-violating term \(\mathcal{O} = Q \text{ or } \mathcal{O}_{W}\).
After the matrix elements have been extracted from the three-point correlation functions
(using fits in the region \(t\gg \tau \gg 0\)), the
CP-odd form factor \(F_{3}(Q^{2})\) can be disentangled from the CP-even form factors
\(F_{1,2}(Q^{2})\) by solving a system of equations.
% Lastly, the expansion relation in (\ref{F3TOEDM}) tells us \(F_{3}(0)/2m = d_{N/P}\). \textbf{We already say this above.}
Since \(F_{3}(0)\) cannot be extracted directly from \(Q^{2}=0\),
an extrapolation to \(Q^{2} \rightarrow 0\) is required. We adopt a linear extrapolation in $Q^2$
following $\chi$PT results~\cite{Ottnad:2009jw, deVries:2010ah, Mereghetti:2010kp}.

In fig.~\ref{TopF3} we show the $Q^2$ dependence of the CP-odd form factors induced by the $\theta$-term.
The data show a signal for the heavy pion mass $m_{\pi} = 701$ MeV while for the light pion mass
$m_{\pi} = 411$ MeV the results are consistent with a vanishing EDM within our statistical uncertainties.
This results is not surprising because the $\theta$-EDM vanishes in the chiral limit in the continuum theory.
The results from the proton and the neutron EDM differ in sign as well as the slope in $Q^2$.
This is consistent with what is expected from $\chi$PT~\cite{Ottnad:2009jw, deVries:2010ah, Mereghetti:2010kp}.
\begin{figure}
  \vspace{-1cm}
  \includegraphics[trim={0cm 0cm 0.5cm 0cm},width=0.5\textwidth]{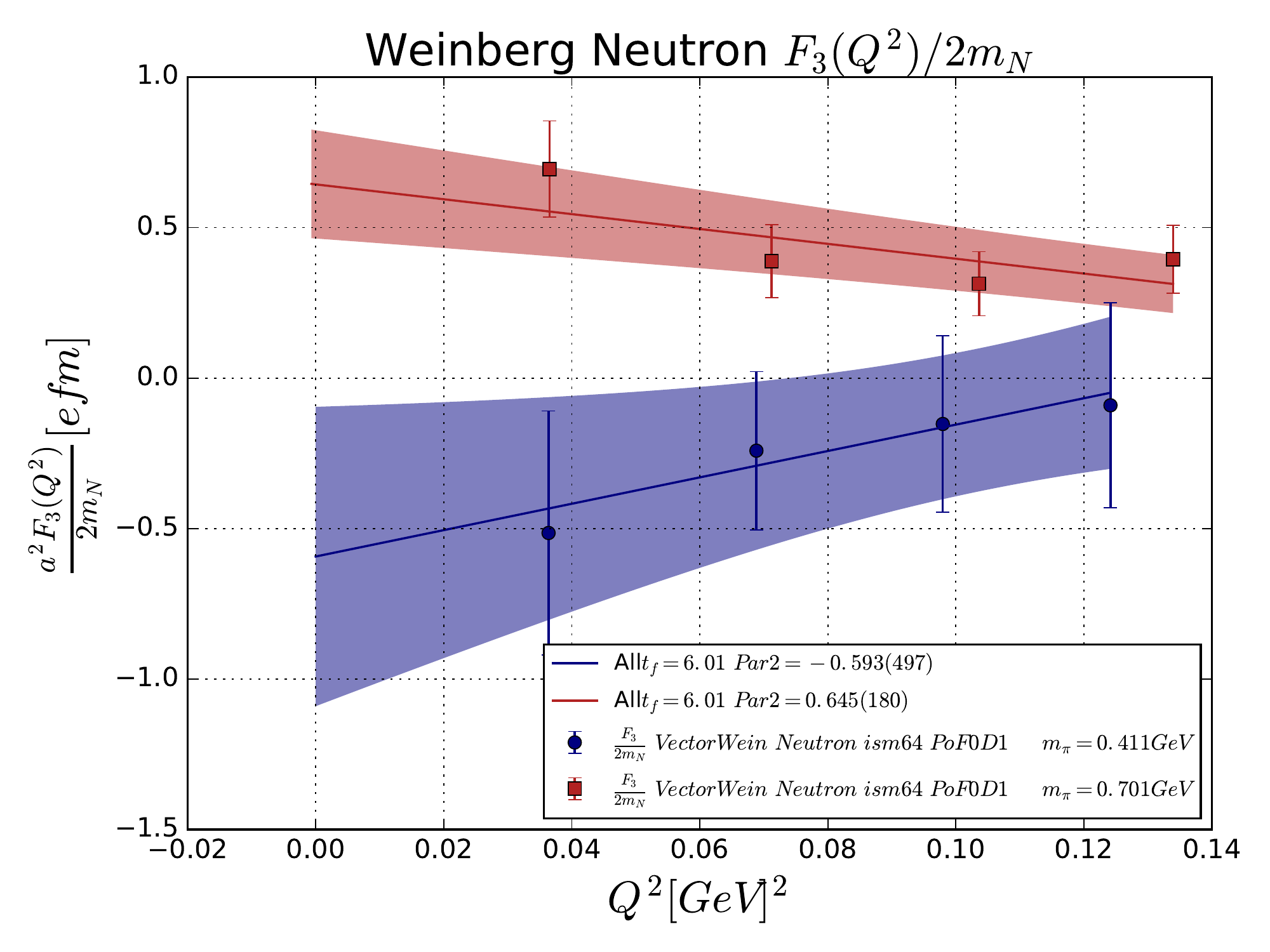}
  \includegraphics[trim={0cm 0cm 0.5cm 0cm},width=0.5\textwidth]{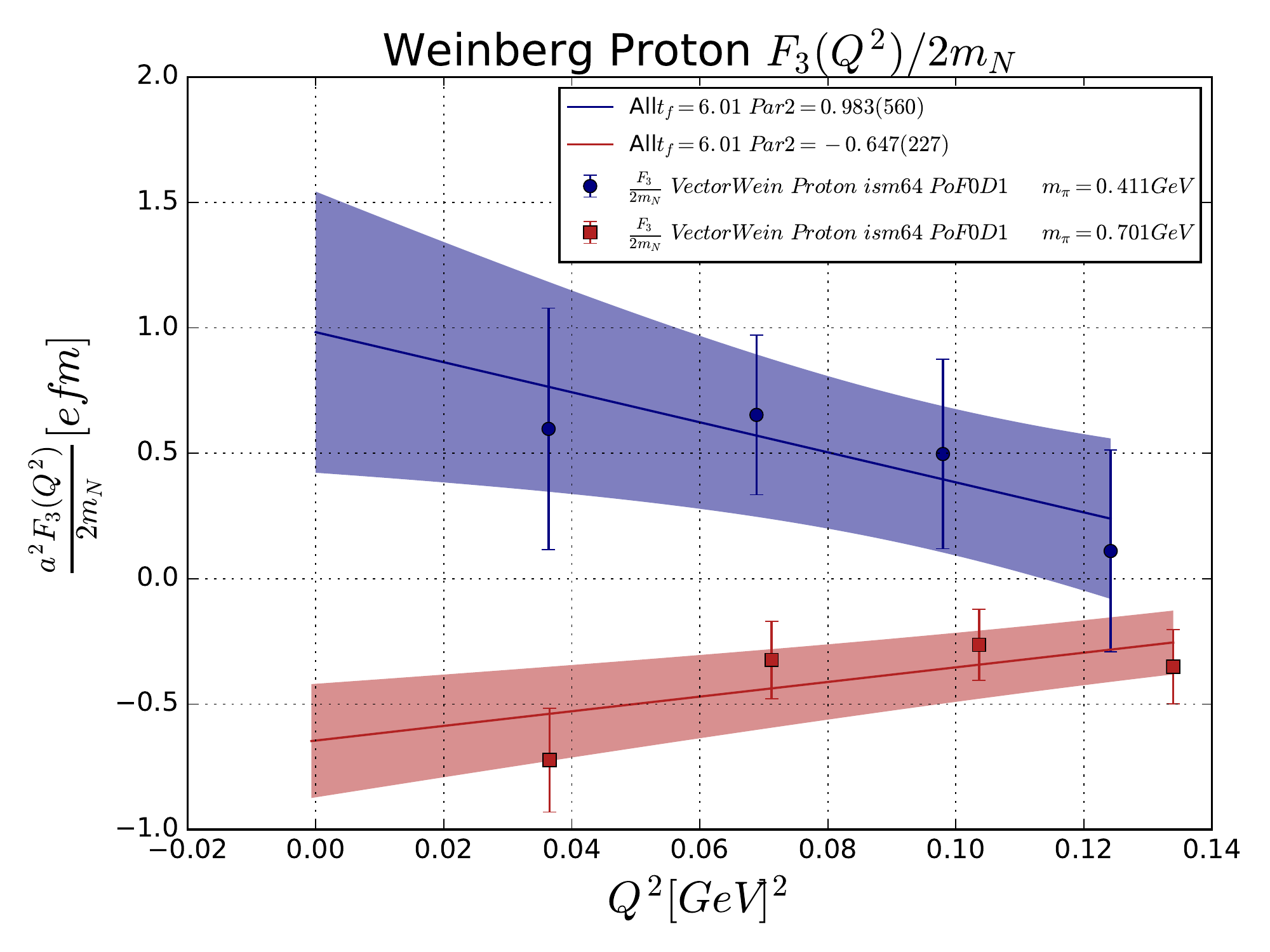}
  \caption{\label{WeinF3} CP-odd vector form factor \(F_{3}/2m_N\) in units of \(\alpha_{W}/\Lambda^{2}\) for the neutron (left) and proton (right),
    induced via the \textit{Weinberg operator} CP-violating vacuum,
    plotted against transfer momentum \(Q^{2}\).
    Blue and red points are the \(m_{\pi} = 411, 701\) MeV results respectively.
    The linear fit to \(Q^{2}\rightarrow 0\) is used to extract the value for the neutron and proton EDM \(d_{n/p}\).}
\end{figure}
%The result for the neutron EDM coming from the \(\theta\)-term contribution
%can estimate from the left plot in fig.\ref{TopF3} by the bound put on by the blue \(m_{\pi} = 411MeV\) results.
%This number is of the order of \( \abs{d^{(\theta)}_{N}} < 0.015\ \theta\ fm\).
%Although the red \(m_{\pi} = 701MeV\) show a clear signal,
%it only helps constrain the chiral extrapolation to be bounded by the lighter blue results.
%In comparison to the experimentally observed \(|d_{n}| < 3.0 \times 10^{-13} \)~e~fm \cite{Olive:2016xmw},
%where \cite{Ottnad:2009jw} uses chiral perturbation theory to come to the conclusion
%that \(\abs{d_{N}} < 0.0012\ \theta\ fm\),
%which is an order of magnitude more precise than this \(\theta\)-term result.
%Similarly, the result for the proton EDM coming from the \(\theta\)-term
%term contribution can estimate from the right plot in fig.\ref{TopF3} by the bound put on by the blue \(m_{\pi} = 411MeV\) results. This number is of the order of \( \abs{d^{(\theta)}_{N}} < 0.02\ \theta\ fm\), slightly less bounded than the proton. It is highly likely that his would be due to systematic effects from the lattice QCD calculation, or even possibly disconnected quark loop contributions which have not been included in this determination.
In fig.~\ref{WeinF3} we show the same results for the CP-odd form factor induced by the Weinberg operator.
Here $\chi$PT is less of a guide for the $Q^2$ and mass extrapolation because pion loops only enter at higher order \cite{deVries:2010ah}.
As a first attempt we still perform a simple linear extrapolation in $Q^2$.
The data indicate that the EDM induced by the Weinberg operator changes sign when lowering the pion mass.
%something not expected from $\chi$PT.
As for the $\theta$-EDM the proton and neutron EDM have opposite sign as well as the slope in $Q^2$.
\begin{figure}
    \vspace{-0.5cm}
  \includegraphics[trim={0cm 0cm 0.5cm 0cm},width=0.5\textwidth]{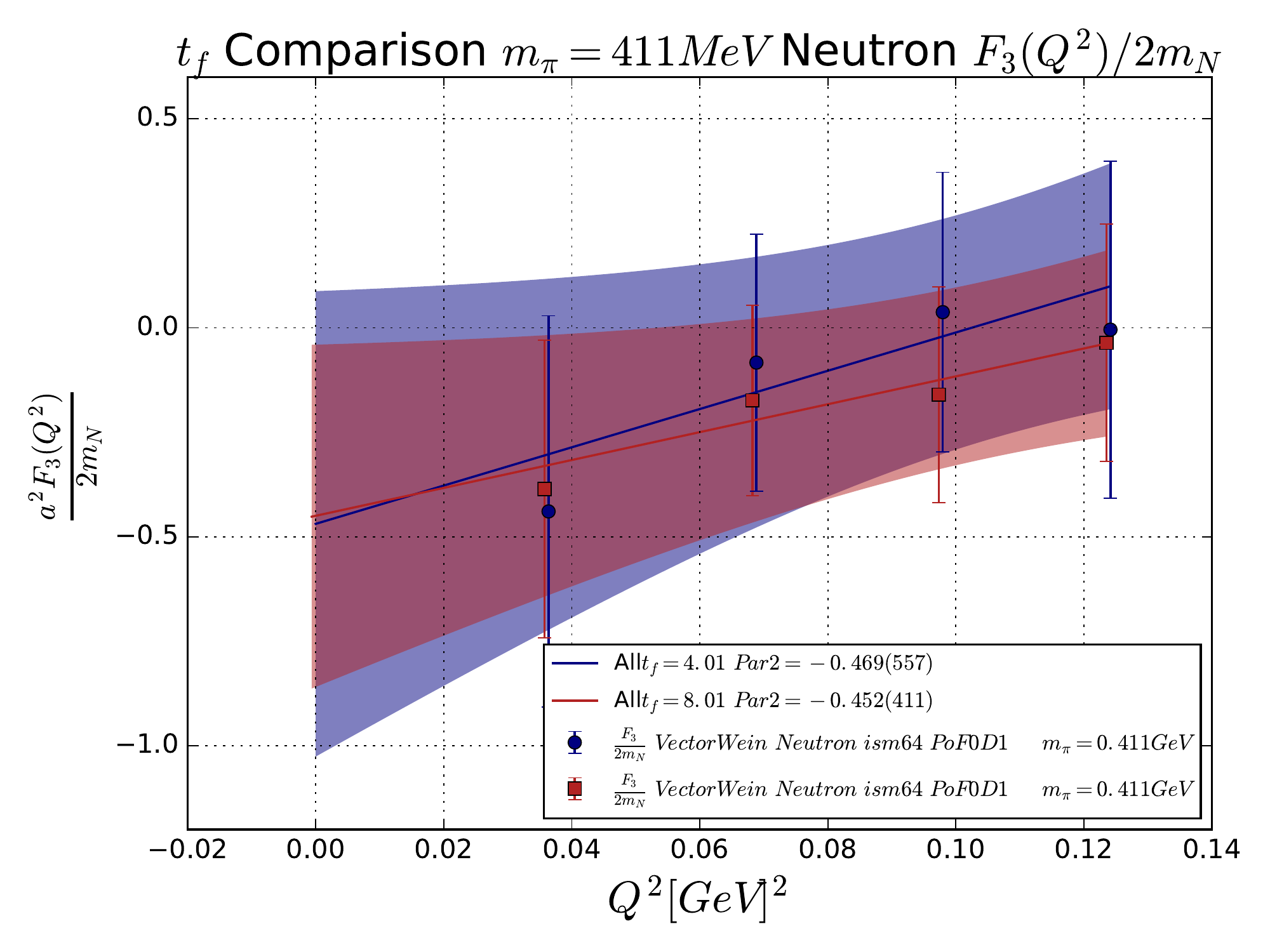}
  \includegraphics[trim={0cm 0cm 0.5cm 0cm},width=0.5\textwidth]{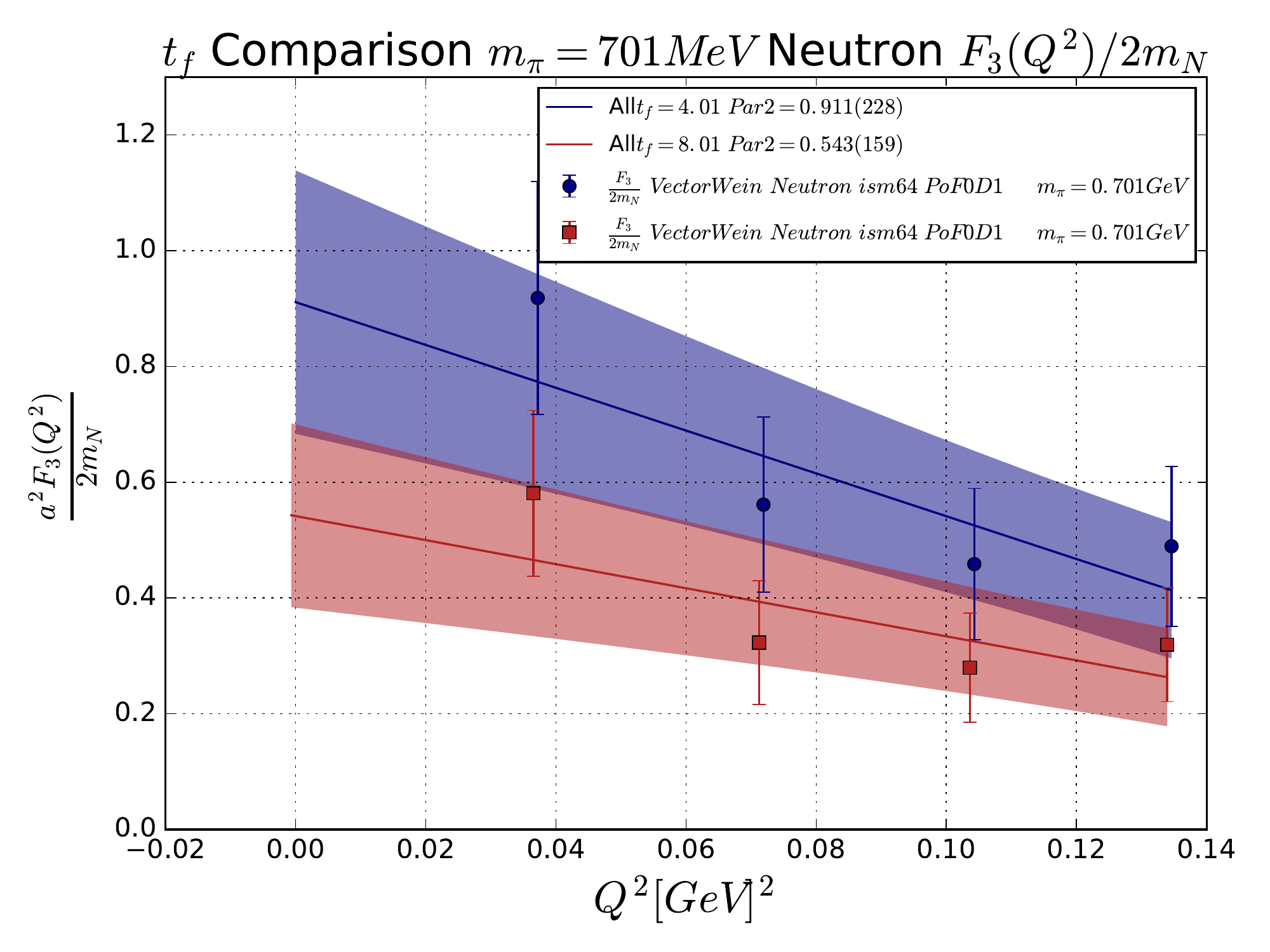}
  \caption{\label{WeinF3FlowDep} Flow time dependence from \(t_{f}/a^2 = 4.01\) (blue) to \(t_{f}/a^2 = 8.01\) (red)
    of the CP-odd vector form factor \(F_{3}/2m_N\) in units of \(\alpha_{W}/\Lambda^{2}\),
    induced via the \textit{Weinberg} CP-violating vacuum, plotted against transfer momentum \(Q^{2}\).
    Left and right plots show the \(m_{\pi} = 411, 701\) MeV results respectively.
    The linear fit to \(Q^{2}\rightarrow 0\) is used to extract the value for the neutron and proton EDM \(d_{n/p}\).}
\end{figure}

An important aspect of the calculation of the EDM is the evaluation of the CP-odd operators at non-vanishing flow time.
Excluding a region at small ftr, we expect that the $\theta$-EDM is independent on the flow time as it happens
for the topological susceptibility and mixing angle (cfr. secs.~\ref{SuscRes}, \ref{NucMixAng}). We have verified that for the EDM
in the range $4 \le t_f/a^2 \le 8$ the result is independent on the flow time. For the Weinberg EDM we would expect a
non-trivial flow-time dependence stemming from the renormalization properties of the Weinberg operator
(cfr. secs.~\ref{SuscRes}, \ref{NucMixAng}).
The numerical data suggest that for the EDM induced by the Weinberg operator the flow-time dependence
is of the same order of statistical accuracy of our calculation and is a little stronger for the heavier pion mass.
This behavior can be observed in fig~\ref{WeinF3FlowDep} where we show the flow-time dependence of the CP-odd form factors
at 2 different flow times, $t_f/a^2=4.01$ and $8.01$.
%One crosscheck that need to be performed, is the flow time dependence of the Weinberg contribution to the neutron and proton EDM, as for the Weinberg susceptibility (sec.\ref{SuscRes}) we saw strong flow time dependence from the right plot in fig.\ref{ChitRes}.
%% This is expected to soften greatly, as the nucleon mixing angle \(\alpha^{(1)}_{N}\) extraction showed greatly reduced flow time dependence.
%In the left and right plots of fig.\ref{WeinF3FlowDep}, a short (blue points, \(t_{f} = 4.01\)) and long (red points, \(t_{f} = 8.01\)) flow time was selected to demonstrate the flow time dependence for the Weinberg contribution to the neutron EDM.
\begin{figure}
  \vspace{-1cm}
  \includegraphics[trim={0cm 0cm 0.5cm 0cm},width=0.5\textwidth]{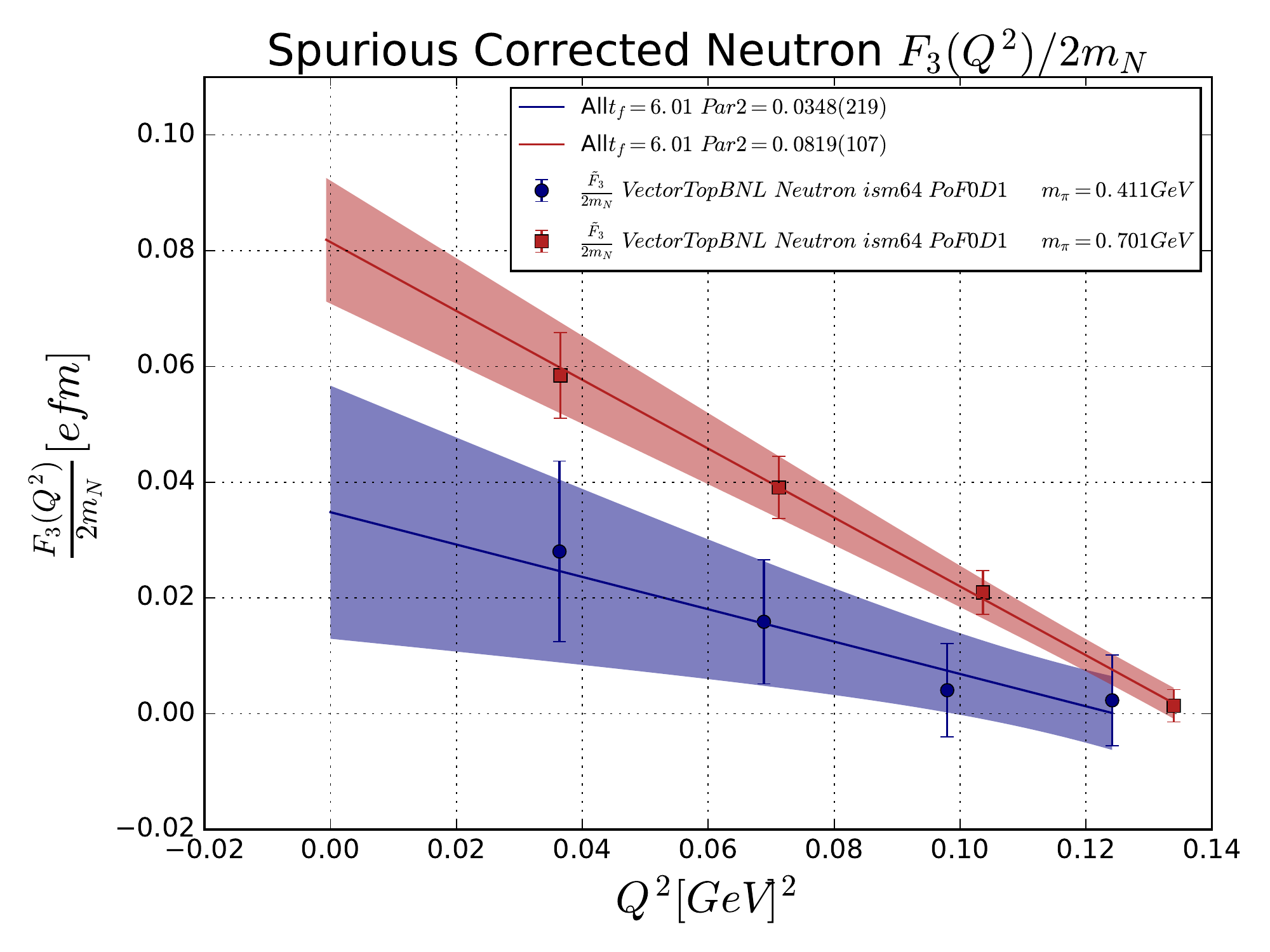}
  \includegraphics[trim={0cm 0cm 0.5cm 0cm},width=0.5\textwidth]{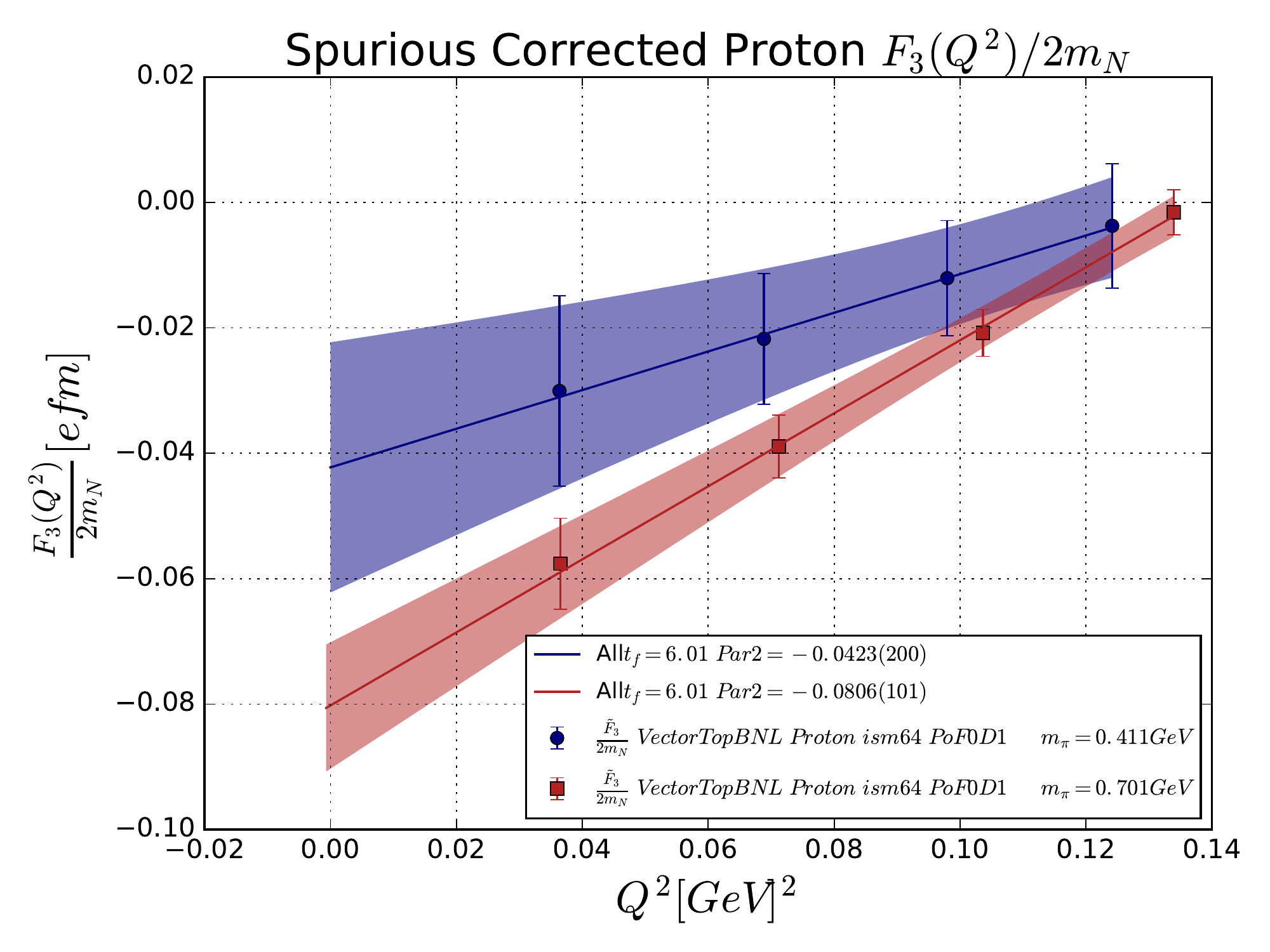}
  \caption{\label{TopF3BNL} Corrections suggested by \cite{Abramczyk:2017oxr} for the CP-odd vector form factor \(F_{3}/2m_N\) in units of \(\theta\) for the neutron (left) and proton (right), induced via the \textit{\(\theta\)-term} CP-violating vacuum, plotted against transfer momentum \(Q^{2}\). Blue and red points are the \(m_{\pi} = 411, 701\) MeV results respectively. The linear fit to \(Q^{2}\rightarrow 0\) is used to extract the value for the neutron and proton EDM \(d_{n/p}\).}
\end{figure}
For the lighter pion mass in the left plot, there is no evidence of flow time dependence,
besides the slight increase in the uncertainty.
Whereas for the heavier pion mass result on the right,
we see a \(1\sigma\) discrepancy between the two extremes of flow time.
\begin{figure}
  \vspace{-0.5cm}
  \includegraphics[trim={0cm 0cm 0.5cm 0cm},width=0.5\textwidth]{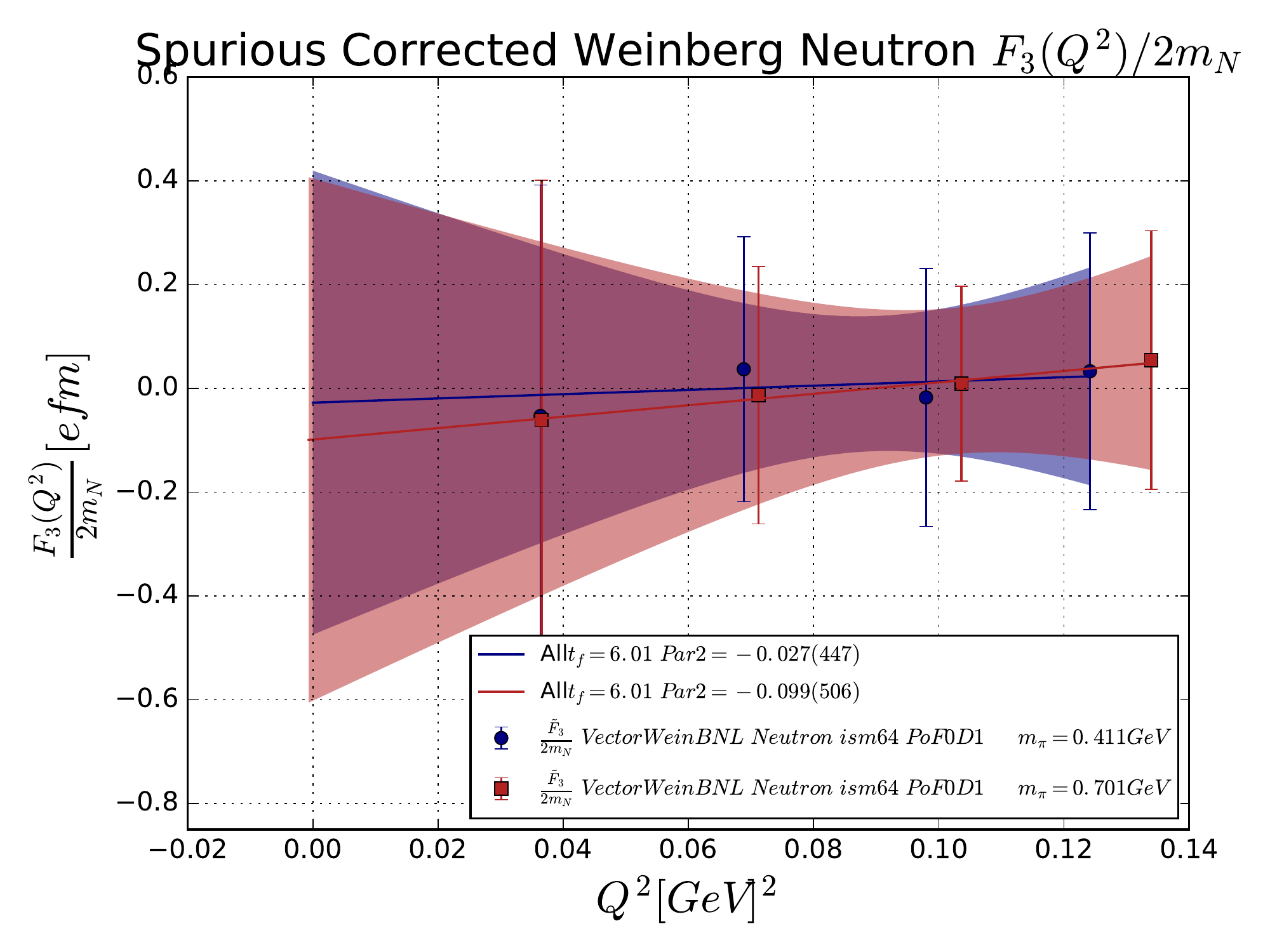}
  \includegraphics[trim={0cm 0cm 0.5cm 0cm},width=0.5\textwidth]{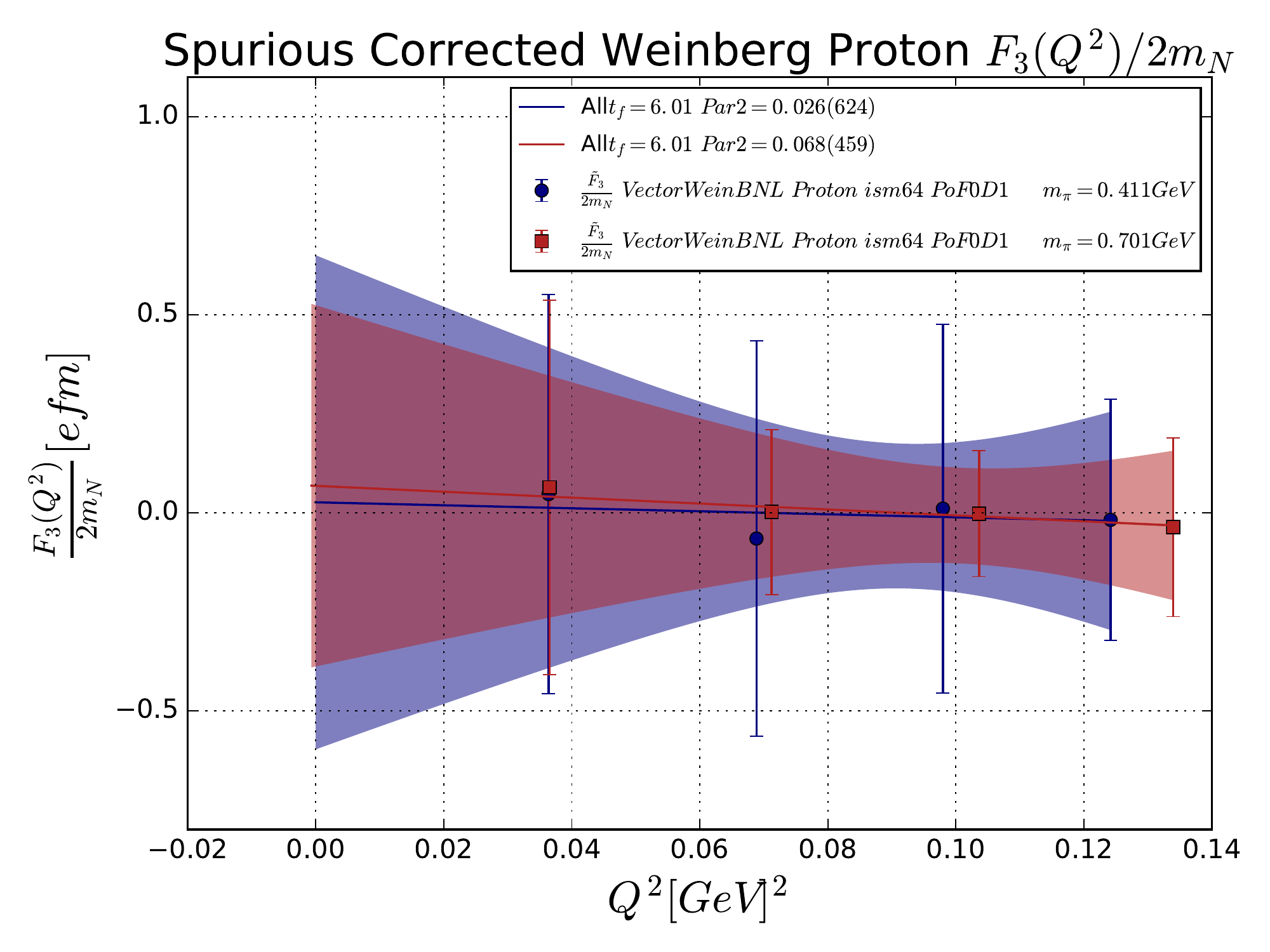}
  \caption{\label{WeinF3BNL} Corrections suggested by \cite{Abramczyk:2017oxr} CP-odd vector form factor \(F_{3}/2m_N\) in units of \(\alpha_{W}/\Lambda^{2}\) for the neutron (left) and proton (right), induced via the \textit{Weinberg operator} CP-violating vacuum, plotted against transfer momentum \(Q^{2}\). Blue and red points are the \(m_{\pi} = 411, 701\) MeV results respectively. The linear fit to \(Q^{2}\rightarrow 0\) is used to extract the value for the neutron and proton EDM \(d_{n/p}\).}
\end{figure}
We have also analyzed the numerical data following ref.~\cite{Abramczyk:2017oxr}. The results are shown in fig.~\ref{WeinF3BNL}.
We observe that the corrections suggested in~\cite{Abramczyk:2017oxr} flip the sign of the $\theta$-EDM (and the slope in $Q^2$)
and make the Weinberg EDM vanish for all pion masses and $Q^2$.
We are currently incresing our statitics and running an additional pion mass to improve our analysis.
%\be
%F_{3} = sin(2\alpha) \tilde{F}_{2} + cos(2\alpha) \tilde{F}_{3}, \qquad
%F_{2} = cos(2\alpha) \tilde{F}_{2} - sin(2\alpha) \tilde{F}_{3},
%\ee
%where the \(\tilde{F}\) signifies the lattice QCD result, and \(F\) is the continuum result.
%The results are presented in figs.\ref{TopF3BNL}\&\ref{WeinF3BNL}.

\subsection{Conclusion}\label{Con}

In this proceeding, we have obtained preliminary results for the $\theta$ and Weinberg EDM
with $N_f=2+1$ dynamical gauge configurations at $2$ pion masses.
We have defined the CP-odd local operators using the gradient flow
for the gauge fields.
We have treated the CP-sources in a perturbative manner allowing us to use existing QCD gauge configurations.
For the $\theta$-EDM we clearly see a signal at the heavier pion mass while the lightest pion mass
is still consistent with zero. Both the sign of the EDMs and the slopes in $Q^2$ of the CP-odd form factors are
consistent with $\chi$PT.
For the Weinberg EDM we see a clear signal at both pion masses, but the observed  pion-mass dependence is not was is expected from $\chi$PT. The signal is washed out after applying
the corrections of ref.~\cite{Abramczyk:2017oxr} and we see no signal at both pion masses.

It is interesting to notice that the flow-time dependence of the $\theta$-EDM is absent in the
large range $4 \le t_f/a^2 \le 8$. We also observe no flow-time dependence for the Weinberg EDM indicating, most likely,
that the variation of the Weinberg operator with the flow time due to its renormalization properties
is below our statistical accuracy. To improve our determination we are currently increasing our statistics
and using a third pion mass in our calculation.

\bibliography{lattice2017}

%%%%%%%%%%%%%%%%%%%%%%%%%%%%%%%%%%%%%%%%%%%%%%%%%%%%%%%%%%%%%%%%%%%%%%%%%%%%%
\end{document}